\documentclass[aps,prb,twocolumn,showpacs,superscriptaddress]{revtex4}
\usepackage{epsfig,psfig,pslatex,latexsym,times,amssymb,amsmath,graphicx}
\usepackage[]{revsymb}
\usepackage{bm}
\input{tcilatex}
\begin{document}

\author{M. Prada}
\affiliation{Physics Department, University of Wisconsin-Madison, 1150 University Ave.,
WI 53705, USA}

\author{R. H. Blick}
\affiliation{Electric and Computer Engineering, University of Wisconsin-Madison, 1415
Engineering Drive, Madison, WI 53705, USA}

\author{R. Joynt}
\affiliation{Physics Department, University of Wisconsin-Madison, 1150 University Ave.,
WI 53705, USA}
\date{\today}

\title{Singlet-Triplet Relaxation in Two-electron Silicon Quantum Dots}

\begin{abstract}
We investigate the singlet-triplet relaxation process of a two electron
silicon quantum dot. \ In the absence of a perpendicular magnetic field, we
find that spin-orbit coupling is not the main source of singlet-triplet
relaxation. \ Relaxation in this regime occurs mainly via virtual states and
is due to nuclear hyperfine coupling. In the presence of an external
magnetic field perpendicular to the plane of the dot, the spin-orbit
coupling is important and virtual states are not required. \ We find that
there can be strong anisotropy for different field directions: parallel
magnetic field can increase substantially the relaxation time due to Zeeman
splitting, but when the magnetic field is applied perpendicular to the
plane, the enhancement of the spin-orbit effect shortens the relaxation
time. \ We find the relaxation to be orders of magnitude longer than for
GaAs quantum dots, due to weaker hyperfine and spin-orbit effects.
\end{abstract}

\pacs{72.25.Rb, 03.67.Pp, 68.65.Hb, 85.35.Be}
\maketitle


\section{Introduction}

A promising technology for the implementation of quantum computation (QC)
involves the storage of quantum information in the spin of electrons in
quantum dots (QDs). \ The key requirement is that the lifetime of the spins
is long compared with the time required for the operation of logic gates. \
This has motivated the development of dots in silicon, where spin-orbit
coupling is weak and isotopic enrichment can eliminate hyperfine coupling
(HC). \ Indeed, recent experiments demonstrate the presence of long-lived
spin states in silicon QDs \cite{nakul}. \ Understanding the processes that
relax spins can point to strategies for minimizing relaxation and coherence
times, thereby improving coherent control of quantum systems. \ \ In the
case of electron spins embedded in semiconductor nanostructures, the
relaxation properties are strongly affected by the regime of operation. Thus
it is important to identify the dominant sources of fluctuations in these
systems, the mechanisms by which they couple to the spins, and to analyze
the non-equilibrium decay laws in different regimes of external fields. \
Considerable theoretical work has been performed on lifetimes for
single-electron spin flip $T_{1}$ and dephasing $T_{2}$ for GaAs \cite%
{khaetskii1,woods,golovach} and for Si \cite{charlie}. \ In GaAs these times
have been measured. \ Single-spin values for $T_{1}$ of about 0.5 ms at a
field of 10 T up to 170 ms at 1.75 T were obtained \cite{elzerman, amasha},
while for $T_{2}$ one finds a value of about 1 $\mu s$ \cite{petta.sci.05}.
\ A qubit consisting of the singlet and triplet states of a two-electron
system is also a proposal for QC. \ \ The singlet-triplet lifetime has been
studied in GaAs \cite{fujisawa,hanson,sasaki,meunier}. \ In particular,
Hanson et al. found $T_{1}$ for the singlet-triplet transition in a
two-electron GaAs dot to be 2.6 ms at B=0.02 T. \ We shall call this $T_{%
\mathrm{ST}}.$ \ Extensive theoretical work has been done for $T_{\mathrm{ST}%
}$ in GaAs \cite{dickmann,florescu,chaney,climente,golovach} and our methods
are similar to those found in these references. \ \ 

In this paper we study the relaxation process for a doubly-occupied Si QD in
an excited (triplet) state to the ground (singlet) state, focusing on the
computation of $T_{\mathrm{ST}}$. \ Our main motivation is to understand
transport through double quantum dots. \ Thus we are concerned with lateral
dots defined by gates in strained silicon quantum wells. \ Such dots have \
a two-fold valley degeneracy as well as spin degeneracy, but we will deal
here with dots where the valley splitting is large compared with the first
orbital excitation energy. \ We will focus on natural Si with a 4\%
concentration of $^{29}$Si nuclei, since this is the material on which
experiments have been performed, but we comment on isotopically enriched Si
below. \ \ 

\begin{figure}[tbh]
\centering\includegraphics[angle=0, width = 0.45\textwidth]{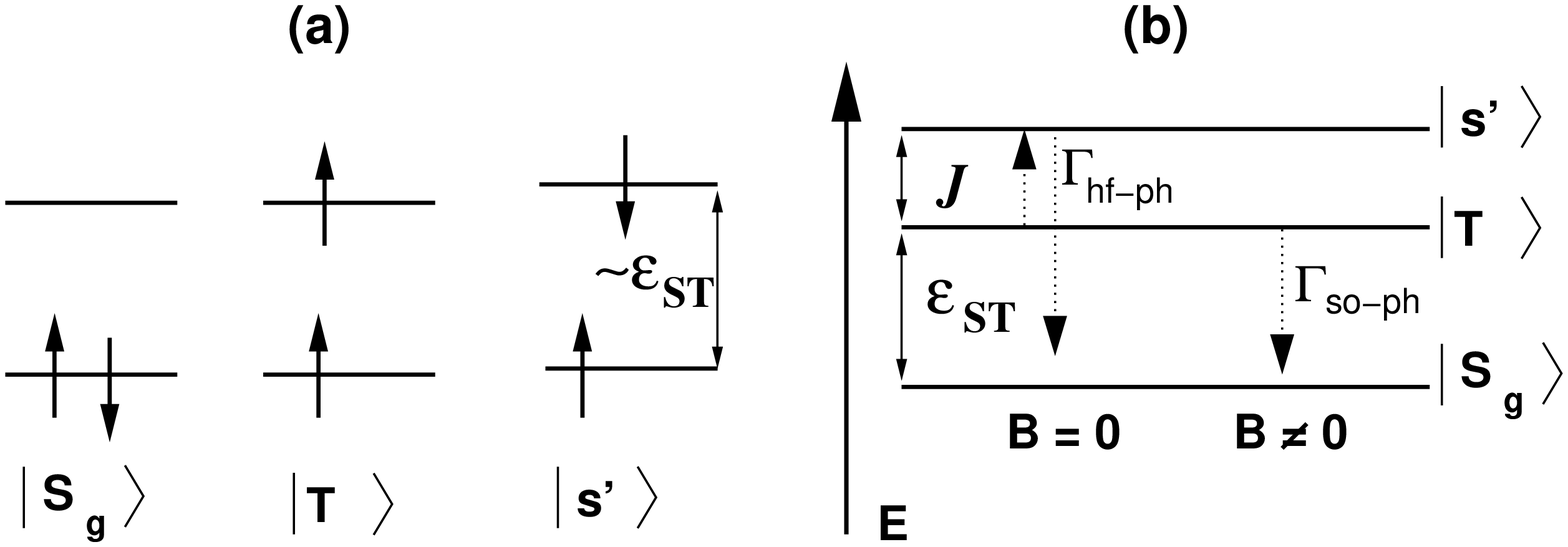}
\caption{\textit{{\protect\footnotesize (a) Level scheme for a
doubly-occupied silicon QD: the ground singlet involves only one orbital,
whereas an additional orbital is needed to form the triplet and higher
states, increasing the level separation. \ (b) Dominant processes in the
quantum dot. \ The two rates indicated are the combined hyperfine and phonon
rate, $\Gamma _{\mathrm{hc-ph}}$ and the combined spin-orbit and phonon
rate, $\Gamma _{\mathrm{so-ph}}$. \ The energy separation of the
ground-state singlet and first triplet is denoted by $\protect\epsilon _{%
\mathrm{ST}}=E_{\mathrm{T}}-E_{\mathrm{S_{g}}}$, and the exchange splitting
by $J$. }}}
\label{fig:levels}
\end{figure}

We assume the levels to be ordered as shown in Fig. \ref{fig:levels}: The
relevant energy scales are then the exchange, $J=E_{\mathrm{s^{\prime }}}-E_{%
\mathrm{T}}$, and the difference between the ground singlet and the first
triplet, $\epsilon _{\mathrm{ST}}=E_{\mathrm{T}}-E_{\mathrm{S_{g}}}$, where
the triplet is formed with a higher energy orbital, as depicted in Fig. \ref%
{fig:levels}(a). \ The dominant mechanism available in the absence of an
external magnetic field is the hyperfine coupling with nuclei \cite%
{saykin,khaetskii} via a virtual state \cite{falko} (left arrows of Fig. \ref%
{fig:levels} b). \ HC cannot cause a direct $T\rightarrow S$ transition
because the nuclei cannot absorb the energy. \ So the change in energy of
the electron spin must be accompanied by the emission of a phonon \cite%
{falko,nazarov,abalmassov}. The electron-phonon interaction mixes thus
different orbital states via a deformation potential in this process, while
the spin-flip is provided by the HC. \ This is the dominant process at zero
applied magnetic field. \ A second relaxation channel is through spin-orbit
(SO) coupling. \ SO coupling mixes different spin states through the \emph{%
Rashba} SO coupling \cite{sobook}. This leads to a non-vanishing matrix
element for the phonon-assisted transition between a singlet and a triplet
state {\textit{in the absence of time-reversal symmetry}} (right of Fig. \ref%
{fig:levels}) leading to an increase in the relaxation rate, $\Gamma _{%
\mathrm{ST}}$ as the field is increased. \ Our aim here is to compute the
singlet-triplet relaxation rates due to these two mechanisms as a function
of external field. \ \ 

We outline the method in the following section, and justify the
approximations that are made. \ We then present results and discussion.\ 
%

\section{Method and Results}

We consider first the case of low fields. \ For reasons to be discussed
below, we may neglect the spin-orbit coupling and the Hamiltonian is written
as $H_{0}+\delta H$, where $H_{0}$ contains the confining potential of the
QD and $\delta H=H_{\mathrm{hc}}+H_{\mathrm{ph}}$. \ Here $H_{\mathrm{ph}}$
corresponds to the electron-phonon coupling (which conserves spin), and $H_{%
\mathrm{hc}}$ is the hyperfine coupling that causes spin mixing in the dot.
\ The confining potential is taken as parabolic with circular symmetry. \ We
do not consider the Coulomb interaction explicitly in the Hamiltonian, as
was done, for example, by Golovach \textit{et al. }\cite{golovach}, whose
main interest was in the regime close to the singlet-triplet crossing. \
Instead the interaction is included phenomenologically through a parameter $J
$, the singlet-triplet splitting. \ This is a crude form of mean field
theory, but is reasonable as long as we are far from the singlet-triplet
crossing point. \ Far from this point there are no energy denominators that
depend sensitively on the interaction strength, and matrix elements depend
smoothly on the strength.

\ For the purpose of this paper (QD formed in a bi-axially strained quantum
well grown along the $z$-axis), we consider the lowest electric subband, so
that the wavefunction $\chi (z)$ in the $z$-direction is fixed. \ We utilize
the Fock-Darwin (FD) states for the lateral dimensions $\phi ^{(n,m)}\left(
x,y\right) $ to construct our wavefunctions, $\{\Psi _{i}\}$, that
diagonalize $H_{0}$. \ For these states, orbital and spin degrees of freedom
factorize: $|\Psi _{S_{g}}\rangle =\chi (z)\times \lbrack |1\rangle
|1\rangle ]\otimes |S\rangle $, $|\Psi _{T}\rangle =\chi (z)\times \lbrack
|1\rangle |2\rangle -|2\rangle |1\rangle ]/\sqrt{2}\otimes |T\rangle $, and $%
|\Psi {s^{\prime }}\rangle =\chi (z)\times \lbrack |1\rangle |2\rangle
+|2\rangle |1\rangle ]/\sqrt{2}\otimes |S\rangle $. Here, $|S\rangle \equiv
\lbrack |\uparrow \downarrow \rangle -\downarrow \uparrow \rangle ]/\sqrt{2}$
and $|T^{+,0,-}\rangle \equiv \lbrack |\uparrow \uparrow \rangle ;(|\uparrow
\downarrow \rangle +\downarrow \uparrow \rangle )/\sqrt{2};|\downarrow
\downarrow \rangle ;$ denote spin states, $|1\rangle \equiv \phi ^{(0,0)}(r)$
and $|2\rangle \equiv \phi ^{(0,\pm 1)}(r)$, so $|1\rangle |2\rangle \equiv
\phi ^{(0,0)}(r_{1})\phi ^{(0,\pm 1)}(r_{2})$. \ \ The electron-electron
interaction is taken into account only phenomenologically through the
parameter $J.$ \ The amplitude of a transition between the triplet $|\Psi
_{T}\rangle $ and singlet $|\Psi _{S_{g}}\rangle $ ground state via an
excited $|\Psi _{s^{\prime }}\rangle $ state is given in second order
perturbation theory, 
\begin{equation}
\langle \Psi _{S_{g}}|\delta H|\Psi _{T}\rangle \thickapprox \frac{\langle
\Psi _{S_{g}}|H_{\mathrm{ph}}|\Psi _{s^{\prime }}\rangle \langle \Psi
_{s^{\prime }}|H_{\mathrm{hc}}|\Psi _{T}\rangle }{E_{T}-E_{s^{\prime }}}.
\label{eq:st}
\end{equation}%
The transition rate from $|\Psi _{T}\rangle $ to $|\Psi _{S_{g}}\rangle $ is
then given by Fermi's golden rule: 
\begin{equation}
\Gamma _{\mathrm{ST}}=\frac{2\pi }{\hbar }|\langle \Psi _{S_{g}}^{f}|\delta
H|\Psi _{T}^{i}\rangle |_{av}^{2}\delta (E_{i}-E_{f}).  \label{eq:fermi}
\end{equation}%
In this notation, $|\Psi _{T}^{i}\rangle $ denotes the initial state of
electron, nuclei and phonons, $|\Psi _{T}^{i}\rangle \equiv |\Psi
_{T}\rangle \otimes |i_{n}\rangle \otimes |i_{\mathrm{ph}}\rangle $,
likewise $|\Psi _{S_{g}}^{f}\rangle \equiv |\Psi _{S_{g}}\rangle \otimes
|f_{n}\rangle \otimes |f_{\mathrm{ph}}\rangle $. \ The $av$ subscript
indicates that the initial states of the nuclear and phonon systems are
averaged over thermal ensembles, and that the final states of these systems
are summed over. \ In this paper we take the temperature to be 100 mK, as
this is roughly the temperature at which experiments are done. \ The chief
approximations involved in the calculation are the use of second-order
perturbation theory and the truncation of the Hilbert space to just two
singlet states and one triplet state. \ The first approximation is excellent
- the rates turn out to be on the order of seconds; \ at those time scales
the Born-Markov approximation implicit in Golden-Rule calculations is surely
valid - the time scales in the bath are probably of the order of the time
for a phonon to traverse the dot. \ The validity of the second approximation
is less clear - in high-symmetry dots such as we are considering here the
phonons do not couple to highly excited states in the dipole approximation,
but real dots may be more disordered. \ \ 

For silicon under compressive stress along [001], the electron interacts
with a phonon of momentum $\bm{q}$ via deformation potentials \cite{charlie,
herring, hasegawa} so the Hamiltonian reads: 
\begin{eqnarray}
H_{\mathrm{ph}} &=&\sum_{s,\bm q}i[a_{qs}^{\ast }e^{-i{\bm qr}}+a_{qs}e^{i{%
\bm qr}}]q(\Xi _{d}\hat{e}_{x}^{s}\hat{q}_{x}+\Xi _{d}\hat{e}_{y}^{s}\hat{q}%
_{y}+{}  \notag  \label{eq:Heph1} \\
&&(\Xi _{d}+\Xi _{u})\hat{e}_{z}^{s}\hat{q}_{z}),
\end{eqnarray}%
where $\langle n_{q}-1|a_{q}|n_{q}\rangle =\sqrt{(\hbar n_{q}/2M_{c}\omega
_{q})}$, $M_{c}$ is the mass of the unit cell, $n_{q}$ is the phonon
occupation number and $\hbar \omega _{q}$ is the phonon energy. Here $s$
denotes the polarisation of the phonon (two transverse and one
longitudinal), $\bm q$ is the wavevector, and $\Xi _{u}$ and $\Xi _{d}$ are
the electron-phonon coupling parameters. \ This is slightly simpler than the
corresponding Hamiltonian in GaAs because of the absence of the
piezoelectric coupling in (centrosymmetric) Si.

Next, we evaluate the spin-flip matrix element given by $\langle \Psi
_{s^{\prime }}|H_{\mathrm{hc}}|\Psi _{T}\rangle $, which is provided, at low
magnetic fields, by the bath of nuclear spins of the $^{29}$Si isotope \cite%
{saykin}. Accordingly, we consider a contact Hamiltonian, 
\begin{equation}
\widehat{H}_{\mathrm{hc}}=\sum_{i,j}\frac{4\mu _{0}}{3I}\mu _{B}\mu _{I}\eta 
{\bm S}_{i}{\bm I}_{j}\delta (\bm r_{i}-\bm R_{j})=A\sum_{i,j}{\bm S}_{i}{%
\bm I}_{j}\delta (\bm r_{i}-\bm R_{j}),  \label{eq:Hhf}
\end{equation}%
where ${\bm S}_{i}$ (${\bm I}_{j}$) and $\bm r_{i}$ ($\bm R_{j}$)denote the
spin and position of the $i$th electron ($j$th nuclei), and $\eta $ and $A$
are hyperfine coupling constants. Inserting eq. (\ref{eq:Hhf}) and (\ref%
{eq:Heph1}) into (\ref{eq:fermi}), we get an expression for the
singlet-triplet rate: 
\begin{equation}
\Gamma _{\mathrm{ST}}=\Gamma _{\mathrm{ph}}\times \left( \frac{A}{2 J }%
\right) ^{2}\sum_{i,j}[|\langle T|[S_{i}^{+}I_{j}^{-}\delta ({\bm r_{i}}-{%
\bm R_{j}})]|s^{\prime }\rangle |^{2}].  \label{eq:gammast}
\end{equation}%
$\Gamma _{\mathrm{ph}}$ describes the phonon rate derived from (\ref%
{eq:Heph1}), and mixes the different orbitals contained in $|s^{\prime
}\rangle $ and $|S_{g}\rangle $. We use the electric dipole approximation ($%
e^{\pm i\bm{q}\bm{r}}\approx 1\pm i\bm{q}\bm{r}$) in \ref{eq:Heph1}, which
is valid for the range of energies considered here ($\thicksim $ 200$\mu $eV 
\cite{nakul}), 
\begin{eqnarray}
\Gamma _{\mathrm{ph}} &\thickapprox &\frac{(n_{q}+1)}{2\rho _{Si}(2\pi )^{2}}%
\sum_{s}\int d\Omega \int_{0}^{\infty }\frac{dqq^{6}}{\omega _{q}}[\Xi _{d}%
\hat{e}_{x}^{s}\hat{q}_{x}+\Xi _{d}\hat{e}_{y}^{s}\hat{q}_{y}+{}  \notag
\label{eq:gammaph} \\
&&(\Xi _{d}+\Xi _{u})\hat{e}_{z}^{s}\hat{q}_{z})]^{2}[|\langle s^{\prime
}|x|S_{g}\rangle |^{2}(\hat{e}_{x}^{s})^{2}+|\langle s^{\prime
}|y|S_{g}\rangle |^{2}(\hat{e}_{y}^{s})^{2}{}  \notag \\
&&+|\langle s^{\prime }|z|S_{g}\rangle |^{2}(\hat{e}_{z}^{s})^{2}]\delta
(\hbar \omega -(J+\epsilon _{\mathrm{ST}})).
\end{eqnarray}%
To evaluate the integral over momenta of (\ref{eq:gammaph}), we assume an
isotropic phonon spectrum, $E_{\mathrm{ph}}=\hbar \omega _{qs}$ and a linear
dispersion relation, $\omega _{qs}=v_{s}q$, $v_{s}$ being the sound velocity
of the mode $s$.

The sum over $j$ of (\ref{eq:gammast}) can be transformed to an integral by
introducing $C_{n}$ as the volume density of $^{29}$Si nuclei, resulting in
a compact expression for triplet-singlet relaxation: 
\begin{eqnarray}
\Gamma _{\mathrm{ST}} &\thickapprox &\left( \frac{A}{2J}\right)
^{2}C_{n}\left( \int d^{3}R_{j}\left[ |\phi ^{(0,0)}(R_{j})|^{2}-|\phi
^{(0,\pm 1)}(R_{j})|^{2}\right] ^{2}\right) {}  \notag
\label{eq:gammast_final} \\
&&\times \frac{1}{2\rho _{\mathrm{Si}}\hbar }\left( \frac{J+\epsilon _{%
\mathrm{ST}}}{\hbar }\right) ^{5}\sum_{i,s}\frac{\gamma _{si}\langle
x_{i}^{2}\rangle }{v_{s}^{7}},
\end{eqnarray}%
where $\gamma _{si}$ contains the result of the angular integral which
depends on the mode $s$ along the coordinate $i$: $\gamma _{lx}=\gamma
_{ly}=4\pi (\Xi _{d}^{2}/3+2\Xi _{d}\Xi _{u}/15+\Xi _{u}^{2}/35)$, $\gamma
_{lz}=4\pi \left( \Xi _{d}^{2}/3+\Xi _{d}\Xi _{u}/5+\Xi _{u}^{2}/7\right) $, 
$\gamma _{tx}=\gamma _{ty}=4\pi \Xi _{u}^{2}/35,$ and $\gamma _{tz}=4\pi \Xi
_{u}^{2}/70.$

It is important to note that $\Gamma _{\text{ST}}$ is proportional to $%
C_{n}, $ i.e., to the total number of nuclei $N_{n}$ with which the
electrons interact. \ This is consistent with the simple picture that the
relaxation rate is proportional to the mean-square fluctuations in the
random hyperfine field. \ Formulas for spin relaxation rates due to
hyperfine coupling that give an apparent proportionality to $N_{n}^{-1/2}$
are common in the literature, and have given rise to the incorrect notion
that some sort of motional narrowing is at work. \ This is not possible,
since the fluctuations in the nuclear spin system are slow. \ In any case
the rate must vanish as $N_{n}\rightarrow 0.$ \ These formulas are correct,
but they generally involve other parameters that actually vary with $N_{n}.$

\begin{figure}[tbh]
\centering\includegraphics[angle=0, width = 0.3\textwidth]{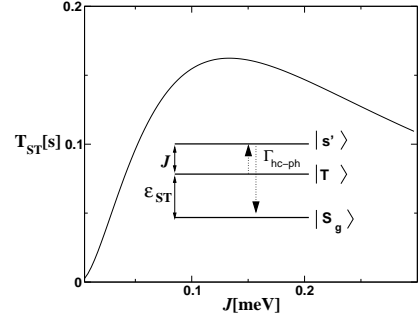}
\caption{\textit{{\protect\footnotesize T$_{\mathrm{ST}}$ as a function of $%
J $ for the process depicted in the inset: when levels $|T\rangle $ and $%
|s^{\prime }\rangle $ are close in energy ($J\rightarrow $ 0), the process
is dominated by hyperfine coupling (fast), and as this energy splits, the
process is dominated by the phonon emission, $~J^{5}$. For higher energies,
the dipole approximation breaks down \protect\cite{cthesis}.}}}
\label{fig:ST}
\end{figure}

Fig. \ref{fig:ST} represents $T_{\mathrm{ST}}=\Gamma _{\mathrm{ST}}^{-1}$
obtained as described in Eq. \ref{eq:gammast_final}, as a function of the
level separation, $J=E_{\mathrm{s^{\prime }}}-E_{\mathrm{T}}$ for a given $%
\epsilon _{\mathrm{ST}}$. In case of Si, $\eta $ = 186, \cite{shulman}
yielding $A\thickapprox 2\times 10^{-7}$eV$\cdot $nm$^{3}$. Only about 4$\%$
of the nuclei have spin, so $C_{n}\thickapprox 0.04\times
8/v_{0}\thickapprox 2$ nm$^{-3}$ ($v_{0}\thickapprox 0.17$ nm$^{3}$). Other
parameters used are: $\Xi _{u}$ = 9.29 eV, $\Xi _{d}$ = -10.7 eV, $\rho _{i}$
= 2330 kgm$^{-3}$, $v_{l}$ = 9330 ms$^{-1}$, $v_{t}$ =5420 ms$^{-1}$, $%
\epsilon _{\mathrm{ST}}$ = 200 $\mu $eV. \ At small $J,$\ $T_{ST}$ increases
as a function of $J:$ the triplet $\left\vert T_{\pm ,0}\right\rangle $and
singlet $\left\vert S^{\prime }\right\rangle $ levels are strongly mixed by
the hyperfine interaction and the phonon density of states increases as a
function of level separations. In the limit $J$=0, the rate is given by
phonon emission, which we found to be of the order of 10$^{12}$ s$^{{-1}}$.
\ Thus $T_{{\mathrm{ST}}}$ appears very small as $J\rightarrow $ 0 in Fig. %
\ref{fig:ST}, but it is not zero. \ At large $J,$ spin mixing is lessened
because there is an energy denominator and phonon relaxation is then
suppressed by spin conservation. \ 

The calculations are obviously consistent with the observed lower bound of \ 
$T_{\mathrm{ST}}>15$ $\mu s$ given in Ref. \cite{nakul}. \ It is expected
that more stringent bounds, hopefully also upper bounds, will be available
soon.   \ 

We now move to the case of finite applied magnetic field $B$. \ We first
take the the field along the growth direction (perpendicular to the 2-DEG): $%
\vec{B}=\vec{B}_{\perp }.$ \ This allows a direct $T\rightarrow S$
transition to occur in the presence of a Rashba SO coupling \cite{charlie1}.
The Rashba field is a consequence of structural inversion asymmetry \cite%
{sobook} in the heterostructure. Note that no bulk inversion asymmetry needs
to be considered in a centrosymmetric crystal like Si. The SO Hamiltonian
due to the full confining potential for the device considered here where $%
l_{\alpha }/r_{\mathrm{QD}}\gg 1$, $l_{\alpha }$ being the Data-Das device
length \cite{datadas}, is then given by \cite{ulloa} 
\begin{equation}
H^{\mathrm{SO}}=\alpha (\vec{\sigma}\times \vec{p})_{z}.
\label{eq:rashba}
\end{equation}%
$H^{\mathrm{SO}}$ mixes the spin states $|S_{g}\rangle $ and $|T^{\pm
}\rangle $, and the orbital wavefunctions as well, so virtual transition to
a higher state $|s^{\prime }\rangle $ is no longer needed and we have: 
\begin{equation}
\langle S_{g}|\delta H_{\mathrm{SO}}|T^{\pm }\rangle \thickapprox \frac{%
\langle i_{\mathrm{ph}}|H_{\mathrm{ph}}|f_{\mathrm{ph}}\rangle \langle
S_{g}|H_{{}}^{\mathrm{SO}}|T^{\pm }\rangle }{\epsilon _{\mathrm{ST}}+\Delta
^{\pm }},\quad \Delta ^{\pm }=\pm g\mu _{B}B.  \label{eq:deltah2}
\end{equation}%
Here, the phonon just ensures energy conservation. Notice that $\langle
S_{g}|\delta H_{\mathrm{SO}}|T\rangle $ is zero if $B=0$. \ This `Van Vleck
cancellation' has been known for decades \cite{abrahams,vanvleck}, but has
been clarified in recent years \cite{halperin} particularly by performing a
spin-dependent unitary transformation in which the first-order term in $%
\alpha $ is eliminated \cite{aleiner}. \ In explicit calculations in the
original basis the cancellation occurs due to the fact that the admixture of 
$T^{+}$ and $T^{-}$ is equivalent in magnitude but opposite in sign. \ The
key point is that spin-orbit-induced transition rates are always
proportional to $B^{2}$ (or higher powers of $B$ in the case of spin 1/2
dots) \cite{charlie}.

Proceeding as in Eqs. (1) to (\ref{eq:gammast}), we find the rate for the
direct transition: 
\begin{equation*}
\Gamma _{\mathrm{ST}}^{\mathrm{D}}=\Gamma _{\mathrm{ph}}^{\mathrm{D}}\times
\left\vert \sum_{i=\pm }\frac{\langle S_{g}|H_{{}}^{\mathrm{SO}%
}|T^{i}\rangle }{\epsilon _{\mathrm{ST}}+\Delta ^{i}}\right\vert ^{2}.
\end{equation*}%
Note that for $B=0$ $\Gamma _{\mathrm{ST}}^{\mathrm{D}}$ vanishes. Since the
phonon does not mix different orbitals in zeroth order in the multipole
expansion to zeroth order (i.e., $e^{\pm i\bm{q}\bm{r}}\approx 1$), $\Gamma
_{\mathrm{ph}}^{\mathrm{D}}$ reads now: 
\begin{eqnarray}
\Gamma _{\mathrm{ph}}^{\mathrm{D}} &\thickapprox &\frac{(n_{q}+1)}{2\rho
_{Si}(2\pi )^{2}}\sum_{s}\int d\Omega \int_{0}^{\infty }\frac{dqq^{4}}{%
\omega _{q}}[\Xi _{d}\hat{e}_{x}^{s}\hat{q}_{x}+\Xi _{d}\hat{e}_{y}^{s}\hat{q%
}_{y}+{}  \notag  \label{eq:gammaphd} \\
&&(\Xi _{d}+\Xi _{u})\hat{e}_{z}^{s}\hat{q}_{z})]^{2}\delta (\hbar \omega
-\epsilon _{\mathrm{ST}})
\end{eqnarray}

Inserting (\ref{eq:deltah2}) into (\ref{eq:fermi}) and evaluating (\ref%
{eq:gammaphd}), we get the spin-orbit-induced relaxation rate, which is
quadratic in $B$ and $\alpha ,$ in agreement with earlier treatments \cite%
{golovach}: 
\begin{equation}
\Gamma _{\mathrm{ST}}^{\mathrm{D}}\thickapprox \left( \frac{4\Delta m\alpha 
}{\hbar ^{3}}\right) ^{2}\langle r^{2}\rangle \sum_{s}\frac{\gamma _{s}}{%
v_{s}^{5}}  \label{eq:gammast2}
\end{equation}%
with $\gamma _{l}=4\pi (\Xi _{d}^{2}+\Xi _{d}\Xi _{u}/2+\Xi _{u}^{2}/5)$ and 
$\gamma _{t}=4\pi (\Xi _{d}^{2}/5+4\Xi _{d}\Xi _{u}/15+2\Xi _{u}^{2}/15)$. \
Here we use $\alpha \thickapprox $ 50 m/s, following Ref. \cite{charlie} and
adapting the result to an electric field of ~10$^{-7}$V/m, an estimate for $%
E_{z}$ for a QD with a 2DEG density of $\thicksim $4$\times $10$^{11}$cm$%
^{-3}$. \ (One should note, however, that this value of $\alpha $ is very
uncertain.) Fig 3 (a) contains in black the relaxation in the absence of the
Rashba coupling ($\alpha =0)$ and in red the relaxation with the additional $%
\Gamma _{\mathrm{ST}}^{\mathrm{D}}$ ($\Gamma _{\mathrm{ST}}+ \Gamma _{%
\mathrm{ST}}^{\mathrm{D}} )$ . \ We plot the results for two 
values of $J.$ \ For $\alpha =0,$ $T_{\mathrm{ST}}$ is a relatively weak
function of $J,$ since the hyperfine interaction is not very sensitive to
field. \ At finite $\alpha ,$ $B_{\perp }$ activates the mixing and $T_{ST}$
decreases rapidly. \ \ \ \ \ 

\begin{figure}[tbh]
\centering\includegraphics[angle=0, width = 0.50\textwidth]{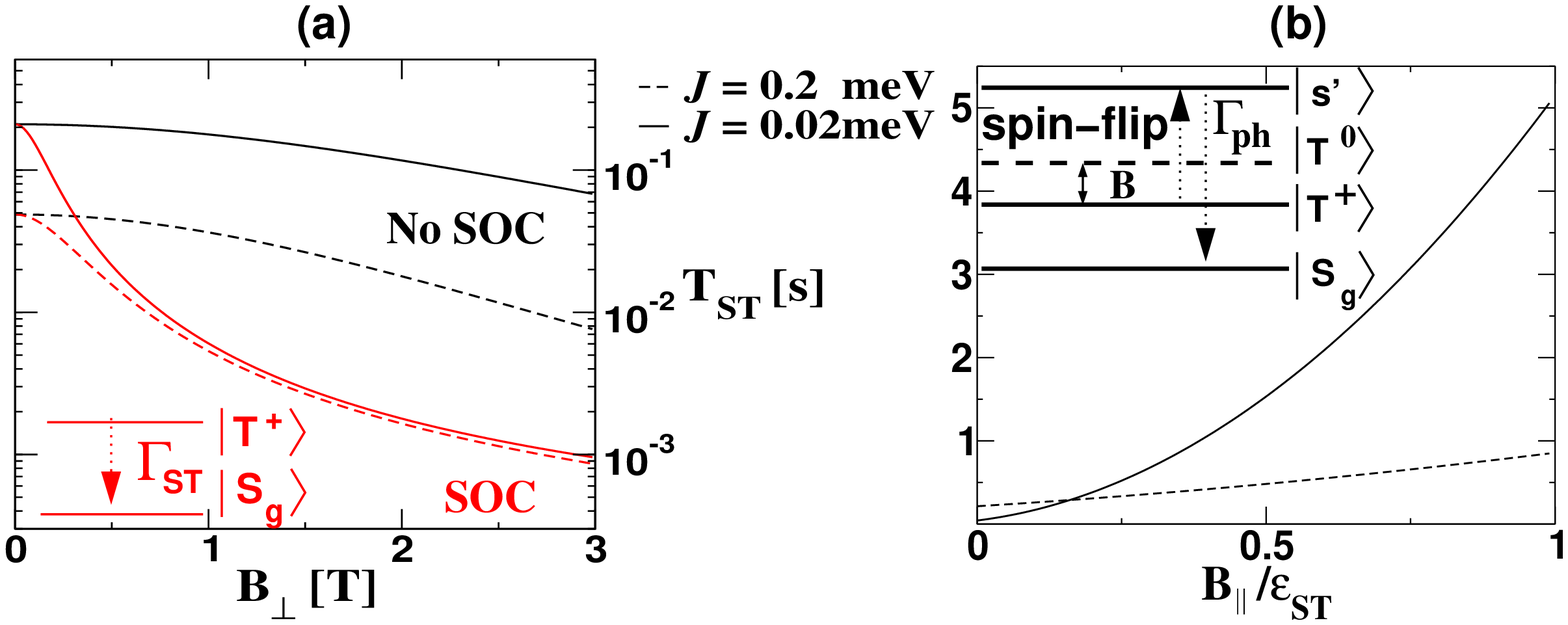}
\caption{\textit{{\protect\footnotesize (Color online) T$_{\mathrm{ST}}$ as
a function of $B $ for two values of $J$: \ The solid lines correspond to a 
plausible value of $J$= 0.02 meV and the broken lines to $J$ = 
$\varepsilon_{\mathrm{ST}}$ =0.2 meV for comparision, (a) as a function of $%
B_{\parallel }$ and (b) as a function of $B_{\perp }$, including Rashba
(red) and without Rashba (black). \ T$_{\mathrm{ST}}$ increases with $%
B_{\parallel }$ and decreases with $B_{\perp }$.}}}
\label{fig:STofB}
\end{figure}

If the magnetic field $B_{\parallel }$ is parallel to the 2DEG, the only
effect is that the spin splitting increases, and larger relaxation times are
obtained (Fig. \ref{fig:STofB}b), as this increases, on average, the energy
separation $E_{|T^{\pm }\rangle }-E_{|s^{\prime }\rangle }$ (see diagram). \
In contrast, the perpendicular magnetic field $B_{\perp }$ decreases the
relaxation time, as we have seen. \ This anisotropy in applied field of $%
T_{ST}$ would be a critical signature of the spin-orbit effect.
We compare our result for a $B_{\parallel }$ = 0.02 T
(T$_{\mathrm{ST}}\thickapprox$ 500 ms) to the experimental value obtained for
GaAs (2.6 ms, \cite{hanson}) and find that Si has a singlet-triplet
relaxation time more than two orders of magnitude larger than GaAs. \ \

Note that the behaviour of $T_{\mathrm{ST}}$ here is different to the much
studied GaAs-based devices \cite{meunier,climente,golovach} because of the
nature of spin-orbit coupling and electron-phonon coupling in
non-centrosymmetric materials: The BIA is absent and there are no
piezo-phonons, also avoided crossings of the singlet and triplet energy
level does not occur for the magnetic fields considered here, giving a
monotonous behaviour.

\section{Discussion}

There are several ways to measure $\Gamma _{\mathrm{ST}}$. \ In single-dot
systems, this can be realized using a single pulse \cite{fujisawa}. \
Alternatively, one may use the following sequence: in the first phase, the
state can be prepared so that only one electron is present in the QD, and in
the next phase, the triplet would be available for conductance, unless it
relaxes to the singlet. \ Measurement of the current for different values of
the pulse duration then gives a direct method to determine $\Gamma _{\mathrm{%
ST}}$. \ The latter experiment has been performed in GaAs \cite{meunier}. \
In double-dot experiments, $\Gamma _{\mathrm{ST}}$ is one of the parameters
in the rate equations that determine the measured current, so these
experiments also provide an avenue for the determination of the
singlet-triplet lifetime \cite{nakul}.

One should note immediately that $T_{\mathrm{ST}}$ is considerably longer in
natural Si than in GaAs, generally by orders of magnitude. \ This is
expected in a system with weaker spin-orbit coupling and fewer spinful
nuclei. \ The times we find are of the order of seconds for the most part. \
It is possible to reduce the time by applying a perpendicular field, which
can serve as a very useful diagnostic. \ It is also possible to lengthen $T_{%
\mathrm{ST}}$ by the use of isotopically enriched Si, i.e., pure $^{28}$Si.%
\ This would eliminate the hyperfine mechanism but it would not get rid of
spin relaxation entirely, as higher-order effects of SOC are still present
even at $B=0$. \ However, these effects are quite small in Si. \ It seems
likely that other effects such as flux noise will be the limiting factor in
isotopically enriched Si.

In summary, we have calculated the dominant rates for phonon-assisted
triplet-singlet relaxation of a silicon quantum dot. \ $T_{\mathrm{ST}}$ is
found to be of the order of hundreds of ms, very sensitive to the exchange
energy $J$, and even longer in the presence of a $B_{\parallel }$, to $~$%
seconds. In the presence of a $B_{\perp }$, a direct transition becomes
possible, increasing (decreasing) $\Gamma _{\mathrm{ST}}$ ($T_{\mathrm{ST}}$%
). \ Due to weak spin-orbit and hyperfine coupling, silicon offers very long
coherence times, which are required for solid state qubits.

We gratefully acknowledge conversations with M. Friesen, S.N. Coppersmith,
A. Vorontsov, M.G. Vavilov and M.A. Eriksson. \ We acknowledge financial
support from the Spanish Ministry of Education and Science (MEC) and from
NSA and ARDA under ARO contract number W911NF-04-1-0389 and from the
National Science Foundation through the ITR (DMR-0325634) and EMT
(CCF-0523675) programs.



\end{document}